\documentclass[journal]{IEEEtran}

\ifCLASSINFOpdf
   \usepackage[pdftex]{graphicx}
\else
  \usepackage[dvips]{graphicx}
\fi

\usepackage{cite}

\begin{document}

\title{Comment on `Interpretation of the Lempel-Ziv Complexity Measure in the context of Biomedical Signal Analysis'}
%
%
%

\author{Karthi~Balasubramanian, Gayathri R Prabhu, Nithin Nagaraj

\thanks{Karthi Balasubramanian and Nithin Nagaraj are with the Department
of Electronics and Communication Engineering, Amrita Vishwa Vidyaapeetham, Amritapuri,
India. e-mail: karthib@am.amrita.edu}}



\maketitle

\begin{abstract}
In this Communication, we express our reservations on some aspects of the interpretation of the Lempel-Ziv Complexity measure (LZ) by Mateo et al. in \cite{LZ_interpretation}. In particular, we comment on the dependence of the LZ complexity measure on number of harmonics, frequency content and amplitude modulation. We disagree with the following statements made in \cite{LZ_interpretation}:
 \begin{itemize}
\item ``LZ is not sensitive to the number of harmonics in periodic signals."
\item ``LZ increases as the frequency of a sinusoid increases."
\item ``Amplitude modulation of a signal doesn't result in an increase in LZ."
\end{itemize}
 We show the dependence of LZ complexity measure on harmonics and amplitude modulation by using a modified version of the synthetic signal that has been used in the original paper. Also, the second statement is a generic statement which is not entirely true. This is true only in the low frequency regime and definitely not true in moderate and high frequency regimes. 

\end{abstract}


\IEEEpeerreviewmaketitle


%

\IEEEPARstart{L}{empel Ziv} complexity has been used in a wide variety of applications including biomedical signal analysis, quantifying regularity of time series and genome data analyis and classification \cite{LZ_interpretation,LZ_radha,LZ_eeg,LZ_eeg_alzh,LZ_dna}. Since LZ complexity is such a popular measure, it is useful to characterize it with respect to various aspects of signal parameters. Mateo et al. in \cite{LZ_interpretation} have presented the use of LZ complexity as a scalar metric for estimating the properties of signals. Some of the experiments and results are reproduced here and we critically review these and point out the fallacies in these results.

\section{LZ Versus Frequency Content}
\begin{figure}[!h]
\begin{center}
\centering
\resizebox{0.9\columnwidth}{!}{
\includegraphics [scale=0.4] {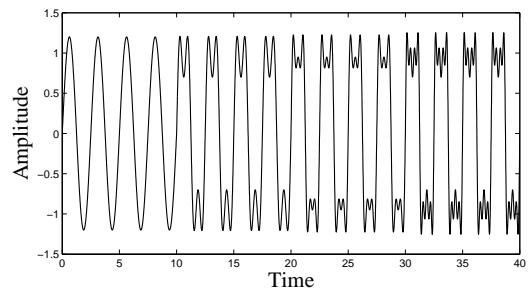}}
\caption {Signal used in \cite{LZ_interpretation} for identifying relationship between LZ complexity measure and frequency content (number of harmonics). }
\label{figure:multitoneSignal}
\end{center}
\end{figure}

In \cite{LZ_interpretation}, the authors propose a test to find out the relationship of LZ complexity value and the frequency content of periodic signals. This test involves concatenation of four periodic signals, each of 10 seconds duration. The first signal was a pure sinusoid and the other three had 3, 5 and 7 frequency components respectively as shown in Fig.~\ref{figure:multitoneSignal}. For this test signal, the LZ complexity value remained a constant throughout and based on this observation, it was concluded that LZ complexity measure is independent of frequency content. But this is an erroneous conclusion.

\begin{figure}[!h]
\begin{center}
\centering
\resizebox{0.9\columnwidth}{!}{
\includegraphics [scale=0.4] {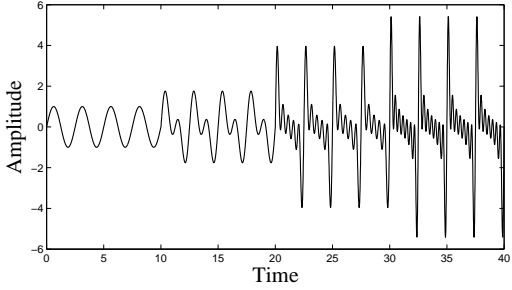}}
\caption {Newly proposed signal with the same frequency content (in terms of harmonics) as the signal in Fig.~\ref{figure:multitoneSignal} in order to determine accurately the relationship between harmonics and LZ complexity measure.}
\label{figure:multitone_correct_Signal}
\end{center}
\end{figure}

It is to be noted that LZ complexity value is calculated after sampling and applying a threshold (\textit{$T_{d}$}) to convert the analog signal in to a 0-1 sequence as given by equation (2) in \cite{LZ_interpretation}. Due to this, the calculated LZ complexity value depends on the signal amplitude variations. For eg: using the median as a threshold, the signal shown in Fig.~\ref{figure:multitoneSignal} will have the same 0-1 sequence as a pure sinusoid existing for the entire 40 seconds. Since the 0-1 sequences are same, the calculated LZ complexity value will also be the same. Thus, the synthetic test signal used for the `LZ Versus Frequency' test in the simulation study in \cite{LZ_interpretation} is incorrect. Having used an erroneous test signal, it is incorrect to interpret that LZ complexity is independent of frequency content since the constant value is due to the limitation of the thresholding mechanism and not due to the calculation of the LZ complexity value. Thus Fig. 2(c) in \cite{LZ_interpretation} is misleading and cannot be used to interpret the dependency of LZ complexity measure on the number of harmonics. This is an artifact due to the limitation of using only two bins for quantizing the data. \\
To be able to correctly model the effect of LZ complexity measure on frequency content, the test signal should be modified such that the generated 0-1 sequence should reflect the changes in the signal. Fig.~\ref{figure:multitone_correct_Signal}  shows one such test signal that we have generated,  that has exactly the same frequency content (in terms of number of harmonics) as the original signal but with amplitude variations required to calculate LZ complexity value accurately.

\begin{figure}[!h]
\begin{center}
\centering
\resizebox{0.9\columnwidth}{!}{
\includegraphics [scale=0.4] {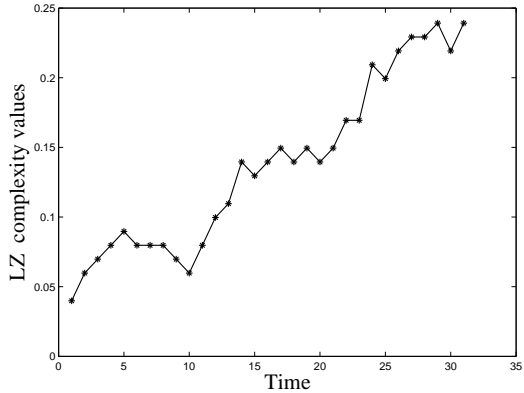}}
\caption {LZ complexity values for the signal shown in Fig.~\ref{figure:multitone_correct_Signal}. It is clear from the above graph that LZ complexity increases with increasing number of harmonics.}
\label{figure:multitone_LZ}
\end{center}
\end{figure}

 Using the Matlab implementation provided by \cite{LZ_calculation} we have plotted the LZ complexity values for the modified signal using a moving window of 10 s with 90\% overlap, similar to what has been done in \cite{LZ_interpretation}. Fig.~\ref{figure:multitone_LZ} shows the resulting plot\footnote{It is to be noted that all plots in this paper show only up to 30 seconds since the sliding window doesn't have enough data after this time interval.}, from which it is evident that the LZ complexity value increases with the increasing frequency content of periodic signals and hence is very much sensitive to the number of harmonics in periodic signals as opposed to what is being claimed in section IV of \cite{LZ_interpretation}. This is important in several biomedical signal analysis applications where it may be needed to identify the number of harmonics in the measured signal. LZ complexity measure is able to capture this information effectively.


In practice, one can use the multi-level LZ complexity measure introduced in \cite{MLZ} to overcome the drawback of having only two bins which loses information.

\section{LZ Versus Frequency}
\begin{figure}[!h]
\begin{center}
\centering
\resizebox{0.9\columnwidth}{!}{
\includegraphics [scale=0.4] {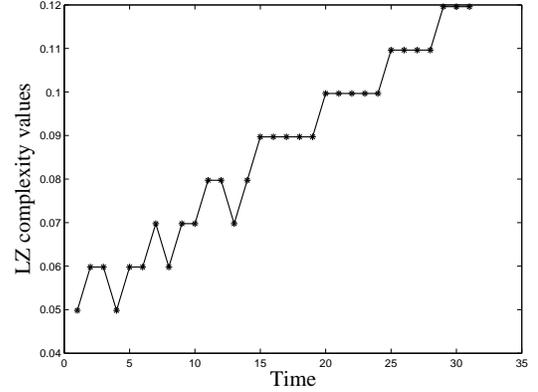}}
\caption {LZ complexity values for chirp signal in low frequency range (0.1 to 1 Hz). The graph indicates the rising trend of LZ complexity with increasing frequencies in this range.}
\label{figure:chirpLowFreq}
\end{center}
\end{figure}

\begin{figure}[!h]
\begin{center}
\centering
\resizebox{0.9\columnwidth}{!}{
\includegraphics [scale=0.4] {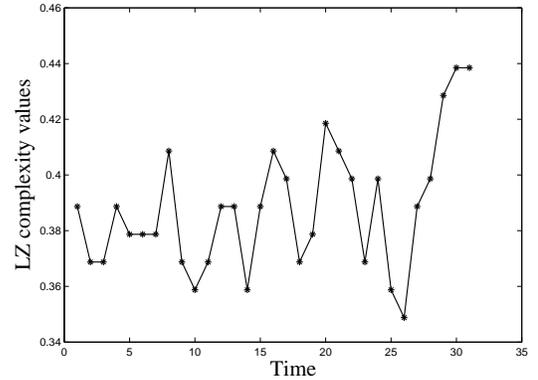}}
\caption {LZ complexity values for chirp signal in high frequency range (5 - 50 Hz). There is no clear trend in LZ complexity with increasing frequencies in this range.}
\label{figure:chirpHighFreq}
\end{center}
\end{figure}

In \cite{LZ_interpretation}, the authors propose to use a chirp signal to determine the effect of frequency on LZ complexity value. Using the results shown in Fig. 2(a) in \cite{LZ_interpretation}, the authors have mentioned that LZ complexity value increases with frequency. But this can't be generalized as the increase in LZ complexity value is seen only in lower frequencies and not in the higher ranges. Fig.~\ref{figure:chirpLowFreq} and Fig.~\ref{figure:chirpHighFreq} clearly indicate the fact that LZ complexity value increases only at very low frequency (0.1 to 1 Hz) and hovers around a constant value at higher frequencies (5 to 50 Hz).

\section{LZ Versus Amplitude modulation}
\begin{figure}[!h]
\begin{center}
\centering
\resizebox{0.9\columnwidth}{!}{
\includegraphics [scale=0.4] {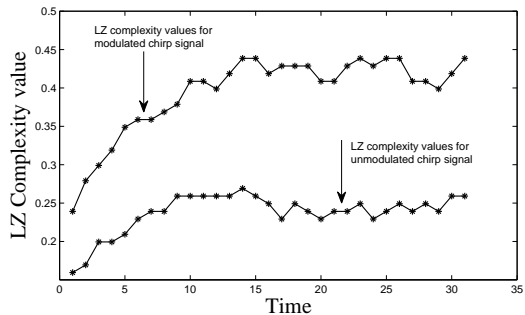}}
\caption {LZ complexity values of chirp signal with and without modulation. The LZ complexity values for modulated chirp signal is clearly higher than the unmoduldated one.}
\label{figure:chirp_nomodulation}
\end{center}
\end{figure}

In order to determine the effect of amplitude modulation on LZ complexity measure, the authors have used an amplitude modulated chirp signal and calculated the LZ complexity value on it and the results are shown in Fig. 2(b) in \cite{LZ_interpretation}. Comparing this with values from an unmodulated chirp signal, it has been incorrectly concluded in \cite{LZ_interpretation} that LZ complexity is independent of amplitude modulation. This is an artifact that is exactly similar to the one presented in Section I. This erroneous interpretation is due to the fact that the amplitude modulated chirp signal and the original chirp signal have exactly the same 0-1 sequence. If the signal is amplitude modulated in such a manner that the generated 0-1 sequence is different, then the changes in LZ complexity values may be noticed. This effect can be seen in Fig.~\ref{figure:chirp_nomodulation} which are plots derived for the same chirp signal but with one of them amplitude modulated by a sinuosid.  The general behaviour of the curves remain the same due to the effect of the chirp signal but the actual LZ complexity values has increased at every step due to the effect of the modulation. Hence it is clear that LZ complexity value changes with amplitude modulation and is not independent of it as incorrectly claimed in \cite{LZ_interpretation}.

\section {conclusion}
In this communication, we point out the fact that LZ complexity measure can also be used as a metric for quantifying the frequency content and the effect of amplitude modulation of a signal, contrary to what was claimed in \cite{LZ_interpretation}. Also the fact that LZ complexity value varies with frequency is applicable only for very low frequencies and can't be generalized for all frequency ranges. Caution needs to be exercised while performing analysis with LZ complexity measure since it depends on:
\begin{itemize}
\item Threshold $T_{d}$ (if we use only 2 bins for quantizing the data - else on the number and size of bins).
\item Sampling Period $T_{s}$.
\item Length of the time series N.
\end{itemize}

A thorough analysis on the exact dependence of LZ complexity on the above parameters is outside the scope of this paper.

\section*{Acknowledgment}

We would like to thank Maneesha Krishnan of Indian Institute of Space Science and Technology, Trivandrum for helping us with literature survey and proofreading the manuscript.

\ifCLASSOPTIONcaptionsoff
  \newpage
\fi

\end{document}